\documentclass[%
longbibliography,
superscriptaddress,
 amsmath,amssymb,
 aps,
]{revtex4}

\usepackage{siunitx}
\usepackage{graphicx}% Include figure files
\usepackage{dcolumn}% Align table columns on decimal point
\usepackage{bm}% bold math
\usepackage{stackrel}
\usepackage{xcolor}

\usepackage{amsmath}
\usepackage{amsfonts}
\usepackage{amssymb}
\newcommand{\1}{1}

\newcommand{\s}{\sigma}

\newcommand{\Gmed}{\Gamma_{\text{med}}}

\newcommand{\beq}{\begin{equation}}
\newcommand{\eeq}{\end{equation}}
\newcommand{\ben}{\begin{eqnarray}}
\newcommand{\een}{\end{eqnarray}}

\newcommand{\Ecoli}{{\it Escherichia coli}}
\newcommand{\ecoli}{{\it E.~coli}}

\newcommand{\bsub}{{\it B.~subtilis}}
\newcommand{\suppFig}{Fig.~S}
\newcommand{\mm}{Methods}

\begin{document}

\title{Universal and idiosyncratic characteristic lengths in bacterial genomes}% Force line breaks with \\

\author{Ivan Junier}
\affiliation{CNRS, TIMC-IMAG, Grenoble, France}
\affiliation{Univ.~Grenoble Alpes, TIMC-IMAG, Grenoble, France}

\author{Paul Fr\'emont}
\affiliation{Center for Interdisciplinary Research in Biology (CIRB), Coll\`ege de France, CNRS, INSERM, PSL Research University, Paris, France}
\author{Olivier Rivoire}
\affiliation{Center for Interdisciplinary Research in Biology (CIRB), Coll\`ege de France, CNRS, INSERM, PSL Research University, Paris, France}

\begin{abstract}
In condensed matter physics, simplified descriptions are obtained by coarse-graining the features of a system at a certain characteristic length, defined as the typical length beyond which some properties are no longer correlated. From a physics standpoint, {\it in vitro} DNA has thus a characteristic length of 300 base pairs (bp), the Kuhn length of the molecule beyond which correlations in its orientations are typically lost. From a biology standpoint, {\it in vivo} DNA has a characteristic length of  1000 bp, the typical length of genes. Since bacteria live in very different physico-chemical conditions and since their genomes lack translational invariance, whether larger, universal characteristic lengths exist is a non-trivial question. Here, we examine this problem by leveraging the large number of fully sequenced genomes available in public databases. By analyzing GC content correlations and the evolutionary conservation of gene contexts (synteny) in hundreds of bacterial chromosomes, we conclude that a fundamental characteristic length around 10-20 kb can be defined. This characteristic length reflects elementary structures involved in the coordination of gene expression, which are present all along the genome of nearly all bacteria. Technically, reaching this conclusion required us to implement methods that are insensitive to the presence of large idiosyncratic genomic features, which may co-exist along these fundamental universal structures.
\end{abstract}

\maketitle

\section{Introduction}

Bacterial genomes display a hierarchy of structures which have been studied from at least four perspectives: (i)~chromosomal conformations, (ii) gene expression, (iii) DNA content and (iv) evolutionary conservation. This diversity of perspectives has revealed a variety of associated lengths, most often observed in just one or a handful of bacteria:

(i) From the perspective of chromosomal conformations, four levels of organization stand out: regulatory DNA loops around $1$ kilobase (1 kb, the typical length of bacterial genes), supercoiling domains around $10$ kb, chromosome interaction domains around $100$ kb and macro-domains around $1000$ kb. Small DNA loops induced by cross-linking of transcription factors contribute to gene regulation~\cite{Matthews:1992tx}. At the next scale, supercoiling domains are segments of chromosomal DNA whose super-helicity is globally affected by the local action of topoisomerases, transcription/replication machineries and nucleoid-associated proteins; atomic force microscopy and genetic studies in \Ecoli~\cite{Postow:2004fp} and {\it Salmonella} strains~\cite{Deng:2005bb} have estimated the length of these domains to~10-20 {kb}. At a larger scale, chromosome conformation capture experiments~\cite{DeWit:2012eo} have revealed interaction domains with enhanced internal interactions between loci~\cite{Le:2013ci}; in {\it Caulobacter crescentus}~\cite{Le:2013ci} and {\it Pseudomonas aeruginosa}~\cite{Lloy:unpub}, these domains extend up to \SI{200}{kb}. At a more global scale, macrodomains spanning up to a quarter of the chromosome specifically localize both along the genome and in the cellular space~\cite{Niki:2000wz,Valens:2004um,Espeli:2008vl}; identified in \Ecoli, they are hypothesized to contribute to chromosome segregation~\cite{Junier:2014ea}.

(ii) From the perspective of transcription, the shortest length of coordination is that of operons~\cite{Jacob:1960ts}, with a mean length (discarding single genes) of \SI{3}{kb} in \ecoli\ and \SI{3.5}{kb} in \bsub, and a maximal length of \SI{18}{kb}~\cite{regDB} and \SI{27}{kb}~\cite{DBTBS}, respectively. Interestingly, the transcription of neighboring genes along the chromosome is significantly correlated beyond operons, up to typically  \SI{20}{kb}~\cite{Jeong:2004jb,Carpentier:2005ws,Marr:2008gg,Jiang:2015bv,Ferrandiz:2016kw,Junier:2016po,Junier:2016cs}. In \ecoli, in addition to being close to the maximal length of operons, this length also corresponds to the maximal length over which functionally related genes are clustered along the chromosome~\cite{Junier:2012dc}. Remarkably, this supra-operonic correlation in transcription is independent of the action of transcription factors and sigma factors~\cite{Junier:2016po}. Instead, it is controlled by DNA supercoiling and transcriptional read-through~\cite{Junier:2016po,Junier:2016cs,MiravetVerde:2017es,Meyer:2017iy}, the capacity of RNA polymerases to transcribe successive operons in one go.

(iii) From the perspective of DNA sequences, several characteristic lengths have emerged from correlation analyses of the GC content of genomes. They include most notably the typical operon length scale below which the GC content of the genome is highly correlated~\cite{BernaolaGalvan:2002uf,BernaolaGalvan:2004gs}. In addition, enhancements of these correlations up to 20 kb have been reported in some bacteria~\cite{BernaolaGalvan:2004gs}. Beyond large differences in overall GC content across bacteria, correlation analyses have also highlighted large intra-genomic variations at multiple scales, which themselves differ between bacteria; the genome of \bsub\ is for instance more heterogeneous than the genome of \ecoli  ~\cite{BernaolaGalvan:2004gs}. These idiosyncratic heterogeneities can cause GC autocorrelation functions to display long tails~\cite{Audit:2003ud}. Besides GC content, clustering genes based on their codon usage has also revealed supra-operonic characteristic lengths~\cite{Bailly:2006PL,Mathelier:2010fm}; the results, again, seem to differ between bacteria, depending on their level of GC heterogeneity: codon usage domains have been shown to extend to $15$ {kb} in the GC-homogeneous genome of \ecoli~\cite{Bailly:2006PL,Mathelier:2010fm} and to \SI{200}{kb} in the GC-heterogeneous genome of \bsub~\cite{Bailly:2006PL}. 

(iv) From the perspective of evolution, some operons, such as ribosomal operons, are extremely conserved. Remarkably, conservation of gene context extends beyond operons~\cite{Lathe:2000um,Tamames:2001wk,Rogozin:2002tia,Fang:2008iv,Junier:2012dc,Junier:2016po} with, in several species, associated lengths reaching 20 kb~\cite{Junier:2016po}, comparable, again, to the maximal lengths of \ecoli\ and \bsub\ operons. Besides synteny (the evolutionary conservation of gene proximity along chromosomes), another signature of evolutionary constraints is the co-occurrence of orthologous genes in different genomes. In closely related strains of \ecoli, it has revealed a primary length of 30 kb associated with functional units and a secondary length of 70 kb reflecting the maximal length of horizontally transferred sequences of DNA~\cite{Pang:2016gk}.

These different observations raise two questions: (1) To what extent lengths associated with distinct properties reflect common underlying structures? (2) To what extent lengths found in a handful of bacteria reflect a universally shared genomic organization? 

To address (1), we previously showed that synteny segments, defined as contiguous sets of genes whose proximity is conserved in a significant number of phylogenetically distant genomes, correspond to domains of conserved co-expression~\cite{Junier:2016po}. Several works strongly suggest that these evolutionary conserved and co-expressed segments can be identified to supercoiling domains~\cite{Lagomarsino:2015kg,Meyer:2017iy}. Connections between the distribution of domains with biased codon usage and gene expression profiles have also been reported in \ecoli~\cite{Mathelier:2010fm}. Yet, to the best of our knowledge, no systematic comparison between lengths associated with sequence correlations (GC content) and other properties has been performed at the scale of the bacterial kingdom. As a result, the question (2) of whether lengths defined in a particular genome reflect an idiosyncratic property of this genome (possibly even confined to a small part of it), or a ubiquitous property shared by most bacterial genomes has remained open.

Here, we address this question by focusing on sequence and evolutionary properties of genomes. Unlike chromosome conformations and gene expression for which experimental data is available only for a few organisms, these properties can be studied across a large number of genomes by leveraging the nucleotide sequences available in public databases. First, we use these datasets to revisit correlation analyses of GC content and show how a fundamental characteristic length can be identified, which is widely shared among bacterial genomes. Second, we analyze distances below which genome organization is evolutionarily conserved and show that a similar characteristic length emerges in nearly all bacterial genomes. Finally, we provide evidence that these two results are the consequence of a common underlying constraint related to the coordination of gene expression.

\section{Results}

\subsection{Characteristic lengths associated with GC content}

\begin{figure}[t]
\begin{center}
\includegraphics[width=.8\linewidth]{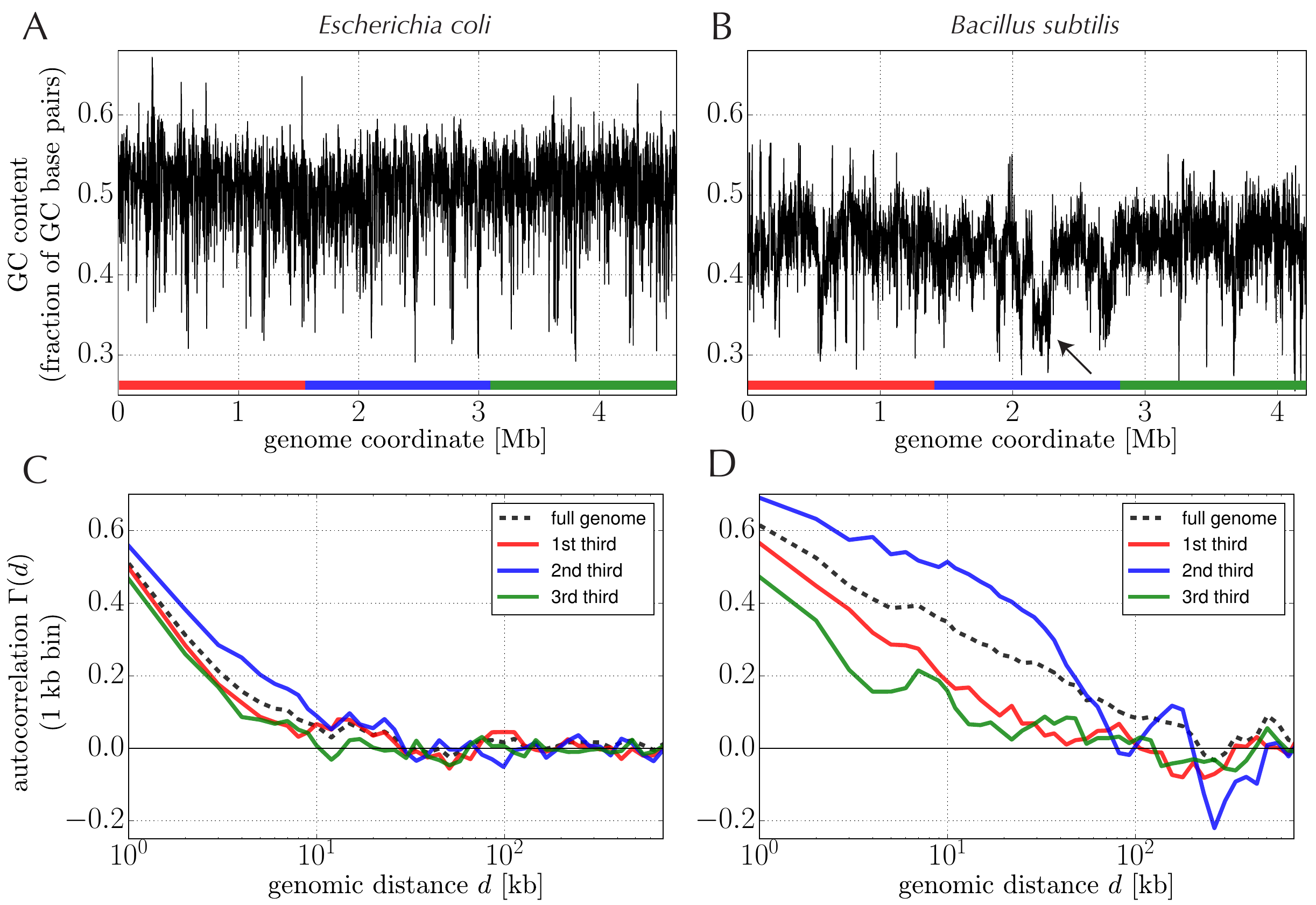}
\caption{Correlations from GC content - {\bf A.} GC content along the chromosome of {\it E. coli}, obtained by computing the fraction of bases that are G or C in successive bins of length $\ell =1$ kb (1000 bases). {\bf B.} GC content along the chromosome of {\it B.~subtilis}, showing a lower mean and more correlated fluctuations than {\it E. coli}; the arrow indicates a fluctuation caused by the presence of the SP-beta prophage. {\bf C.} Autocorrelation function $\Gamma(d)$ of GC content from the full genome of {\it E. coli} (dotted line) and from three distinct subparts of it (colors corresponding to those in A). In each case, correlations are enhanced at short distance, up to $\sim 10$ kb. {\bf D.}~Autocorrelations of GC content for {\it B. subtilis}. Here the full genome displays correlations up to $\sim 200$ kb (dotted line) and large differences are observed between the three distinct subparts of the genome.
\label{fig:1}}
\end{center}
\end{figure}

The GC content of a given sequence of DNA is quantified as the fraction $x$ of its bases that are pyrimidine bases (G and C). As illustrated in Fig.~\ref{fig:1}A-B with the \ecoli\ and \bsub\ genomes, this GC content varies between genomes as well as within genomes. In this figure, intra-genomic variations are represented by partitioning each genome into bins of length $\ell=\SI{1}{kb}$ (the typical length of a gene) and reporting the GC content within each bin.

The characteristic length over which intra-genomic variations are significantly correlated can be read from an autocorrelation function. Given the partition of a genome into $N$ bins of length $\ell$ and given the GC content $x[i]$ within each bin $i$, this function is defined for $d=0,\ell,2\ell,\dots, (N-1)\ell$ as
\beq\label{eq:Gd}
\Gamma (d)=\frac{1}{N\s_x^2}\sum_{i=0}^{N-1} (x[i]-m_x) (x[i+d/\ell]-m_x),
\eeq
where $m_x=\sum_i x[i]/N$ is the mean value of $x$ and $\s^2_x=\sum_i (x[i]-m_x)^2/N$ its variance. As most bacterial genomes are circular, we consider periodic boundary conditions, i.e., $i+d/\ell$ is understood modulo $N$ in Eq.~\eqref{eq:Gd}. The autocorrelation functions $\Gamma (d)$ for \ecoli\ and \bsub\ are represented in a semi-log plot (distances $d$ along the $x$-axis in log scale) as black dotted lines in Fig.~\ref{fig:1}C-D. In the case of  \ecoli, the correlations decrease at short distance, up to around 20~kb, before fluctuating around zero; in the case of \bsub, on the other hand, significant correlations are observed up to around 200 kb.

The GC content of \bsub, however, is quite heterogeneous (Fig.~\ref{fig:1}B). As a consequence, autocorrelation functions computed from different parts of the genome are markably more different than in \ecoli\ (red, blue and green curves in Fig.~\ref{fig:1}C-D, each corresponding to one third of the genome; see \mm\ for details of the computations). Part of the heterogeneity of the \bsub\ genome can be attributed to the presence of prophages, i.e., DNA from bacterial viruses that has been integrated by horizontal transfer~\cite{canchaya2003prophage}. In particular, a 134~kb long prophage known as the SP-beta prophage manifests itself in Fig.~\ref{fig:1}B as an extended region with low GC content (indicated by an arrow); it is mostly responsable in Fig.~\ref{fig:1}D for the larger correlations of the blue autocorrelation function. 

\begin{figure}[t]
\begin{center}
\includegraphics[width=.99\linewidth]{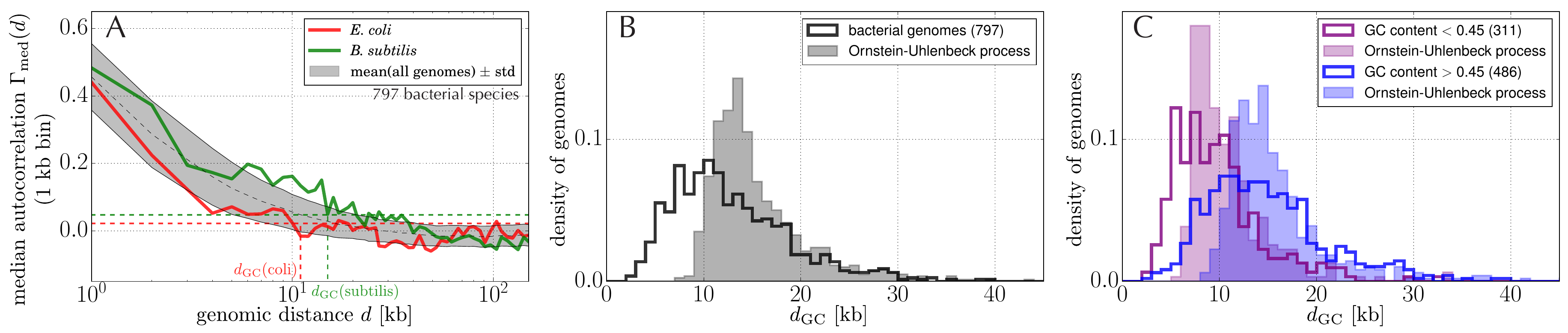}
\caption{Characteristic lengths from GC content -- {\bf A.} Median autocorrelation of GC content for {\it E. coli}~(red),  {\it B. subtilis}~(green) and all genomes (dotted line and gray area representing respectively the mean and standard deviation over 797 genomes). Genomes are partitioned into 500 kb long subsequences, and the median is over the autocorrelation functions computed from each subsequence; this procedure prevents large but isolated heterogeneities in GC content to dominate over smaller but more widespread features. For each genome, a characteristic length $d_{GC}$ is defined by considering the smallest distance $d$ at which $\Gamma(d)$ reaches $\s_\Gamma$, the standard deviation of $\Gamma(d)$ at large lengths $d>100$ kb (red and green dotted lines for the definition of $d_{GC}$ in  {\it E. coli}~and {\it B. subtilis}). {\bf B.} Distribution of characteristic lengths $d_{GC}$ over 797 bacterial chromosomes (empty black histogram). For comparison, distribution of characteristic lengths from a discrete Ornstein-Uhlenbeck process, showing that variations are expected from stochastic fluctuations even in a model with a single correlation length. {\bf C.} Distribution of characteristic lengths $d_{GC}$ for two subsets of the genomes, those with a low mean GC content ($<0.45$, in purple) and those with a high mean GC content ($>0.45$, in blue), showing a tendency for genomes with higher mean GC content to display larger characteristic lengths. As in B, the shaded histograms correspond to expectations from discrete Ornstein-Uhlenbeck processes.
\label{fig:2}}
\end{center}
\end{figure}

This example highlights an important caveat of autocorrelation functions applied to heterogeneous signals that lack translational invariance: the correlations that they report may be dominated by a single or a few localized heterogeneities that conceal more typical correlations associated with most of the signal. But the same example also suggests a workaround: compute the autocorrelation function for distinct subsequences of a genome and, instead of averaging over the resulting autocorrelation functions, consider their median value $\Gamma_{\text{med}}(d)$, which is insensitive to the presence of a few localized heterogeneities. As shown in Fig.~\ref{fig:2}A, where the subsequences result from a partition of the genomes into 500 kb long subsequences, this procedure leaves unchanged the autocorrelation function in the case of \ecoli\ but effectively abolishes the influence of the SP-beta prophage in the case of \bsub. The correlations reported by $\Gamma_{\text{med}}(d)$ are now more similar for \ecoli\ and \bsub. To define for each genome a characteristic length $d_{\text{GC}}$ above which GC content is no longer significantly correlated, we note that $\Gamma_{\text{med}}(d)$ vanishes at large distances and characterize its fluctuations by its standard deviation $\s_{\Gamma}$ for distances $d>\SI{100}{kb}$ (as shown below, these fluctuations are expected from the limited size of genomes). We then define $d_{\text{GC}}$ as the shortest distance at which $\Gamma_{\text{med}}(d)<\s_{\Gamma}$, which yields respectively $d_{\text{GC}}=\SI{11}{kb}$ and $d_{\text{GC}}=\SI{15}{kb}$ for the \ecoli\ and \bsub\ genomes (dotted red and green lines in Fig.~\ref{fig:2}A).

\begin{figure}[t]
\begin{center}
\includegraphics[width=.7\linewidth]{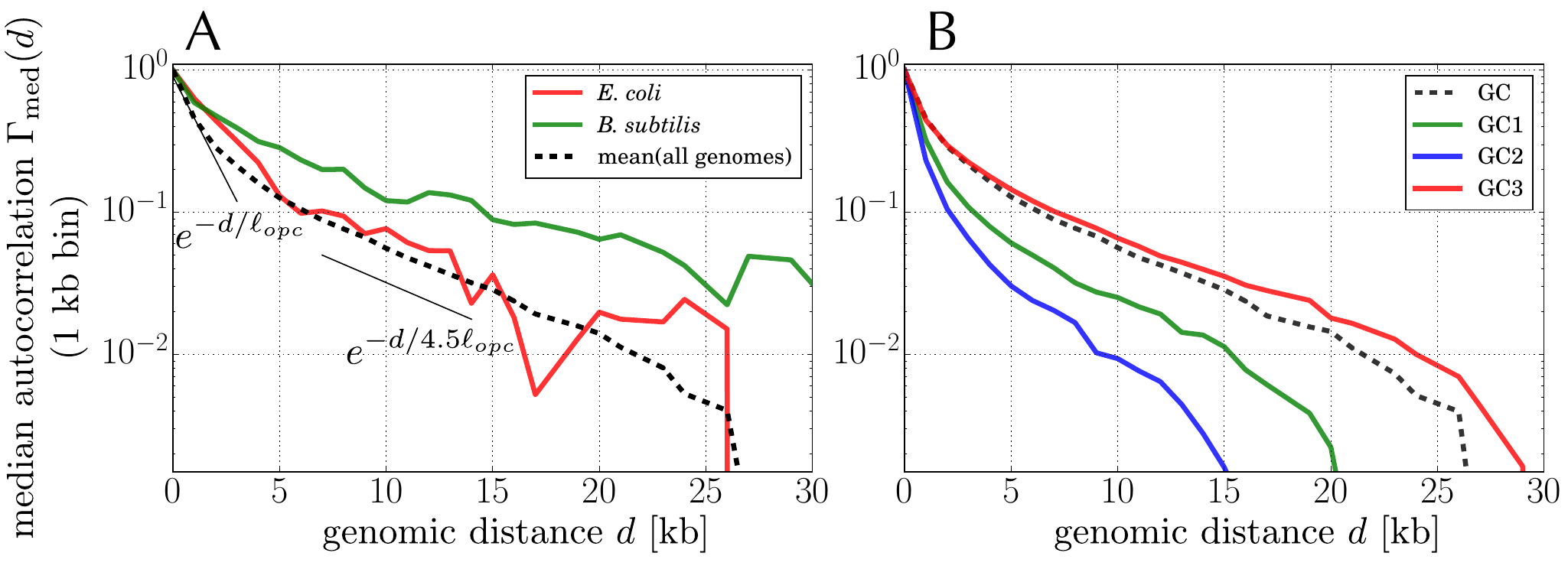}
\caption{Correlation lengths from GC content and codon analysis -- {\bf A.} Median autocorrelation of GC content for {\it E.~coli} (red curve), {\it B.~subtilis} (green curve) and all genomes (black dotted curve). Compared to Fig.~\ref{fig:2}A, the scales along the x- and y-axes are respectively linear and logarithmic; the data is otherwise the same. In this representation, two exponential regimes are apparent as indicated by the thin black lines, each defining a distinct correlation length, the largest one being approximately $4.5$ times larger. The first regime at short distance (below 2-3 kb) is consistent with the exponential distribution of operon lengths  in {\it E.~coli} (including single genes), which is characterized by the same correlation length $\ell_{opc}\simeq \SI{1.5}{kb}$. {\bf B.}~Mean autocorrelation over the 797 genomes of the GC composition computed from the three different bases of gene codons (first base in green, second base in blue and third base in red). For comparison, the autocorrelation over all bases is shown again as a black dotted curve (same curve as in A), showing that it is well recapitulated by the autocorrelation from the third base.
\label{fig:3}}
\end{center}
\end{figure}

Beyond \ecoli\ and \bsub, we analyzed with the same approach a set of 797 chromosomes across the bacterial kingdom. These genomes were selected among a larger set of 1675 genomes from the STRING database~\cite{STRING} on the basis of their suitability for the evolutionary analysis presented below (see \mm). We checked, however, that our results based on GC content hold for the other 861 genomes larger than 500 kb (\suppFig\ref{fig:S1}, second line); we also checked that the results are insensitive to the bin length $\ell$ used for computing GC contents (\suppFig\ref{fig:S1}, first line). Averaged over genomes, the mean autocorrelation function $\langle\Gamma_{\text{med}}(d)\rangle$ has a shape similar to the autocorrelation functions $\Gamma_{\text{med}}(d)$ of \ecoli\ and \bsub\ (Fig.~\ref{fig:2}A). All bacterial genomes indeed display a similar characteristic length $d_{\text{GC}}$ (defined as above; see Fig.~\ref{fig:2}B). More precisely, 91\% of the characteristic lengths lie between 5 kb and 25 kb. Note that as Fig.~\ref{fig:1}C-D,  Fig.~\ref{fig:2}A corresponds to a semi-log plot where the log scale along the $x$-axis represents a wide range of genomic distances. Using a semi-log plot with a log scale along the $y$-axis (Fig.~\ref{fig:3}A) further reveals two underlying correlation lengths. Following the standard definition, we call here correlation length a length  $\zeta$ that characterizes an exponential decay of the form $\exp(-d/\zeta)$ (associated characteristic lengths, above which correlations typically vanish, are usually a few times larger; an example is the Kuhn length, which is twice the correlation length associated with binding properties of DNA, also known as the persistent length~\cite{Strick:2003dp}). These two correlation lengths are manifest in the genomes of \ecoli\ and \bsub\ (red and green curves, respectively, in Fig.~\ref{fig:3}A), and even more clearly in the mean autocorrelation 
(black curve in Fig.~\ref{fig:3}A). Remarkably, the decay of the correlations at short distance (below 2-3 kb) is consistent with the distribution of operon lengths.

To assess the extent to which the variability of characteristic lengths (Fig.~\ref{fig:2}B) arises from statistical fluctuations, we now compare the results to those obtained from a discrete Ornstein-Uhlenbeck (OU) process (also known as the autoregressive model), the archetypal continuous stochastic process with a single correlation length~\cite{karlin1981second}. Our goal here is not to provide a quantitative model of the correlations: as they involve multiple correlation lengths (Fig.~\ref{fig:3}A), they are indeed certainly not described by a simple OU process. Instead, we aim at estimating the variability expected from the finite size of chromosomes, whose total length is typically only 100 times larger than their characteristic length.

Discrete OU signals $\{x[i]\}_{i=0,\dots, N-1}$ are recursively defined by $x[i+1]=ax[i]+\eta_i$ where $a$ is a parameter controlling the correlation length ($-1/\ln(a)$) of the signal and $\eta_i$ is a stochastic term drawn independently for each $i$ from a normal distribution with zero mean and fixed variance $\s^2_\eta$, a second parameter that characterizes the variance of the fluctuations of the signal. In particular, for an infinitely long signal ($N\to\infty$), the signal variance is $\s^2_{\textrm{OU},\infty}=\s^2_\eta/(1-a^2)$, while the autocorrelation function is $\Gamma_{\textrm{OU},\infty}(d)=a^d$. As we are interested in fluctuations of characteristic lengths stemming from finite $N$, we first define $a=0.78$ by fitting $\langle\Gamma_{\text{med}}(d)\rangle$ with $\Gamma_{\textrm{OU},\infty}(d)=a^d$. For each of the 797 genomes, we then generate an OU signal with the number $N$ of bins of length $\ell=1$ kb of the genome, using a common $a=0.78$ and a specific $s_\eta$, obtained by matching $\s_\eta^2/(1-a^2)$ with the variance in GC content of the genome. Finally, we compute for each signal its characteristic length as before (except that we consider $\Gamma(d)$ rather than $\Gamma_{\text{med}}(d)$ as the OU process is free of heterogeneities). The distribution of characteristic lengths over the different genomes obtained by this procedure is shown as a grey-shaded histogram in Fig.~\ref{fig:2}B. The result shows that 65\% (measured as the overlap between histograms) of the variability of characteristic lengths can be imputed to finite-size fluctuations. 

Beyond statistical fluctuations, we find that a systematic source of variability is the overall GC content of a genome: genomes with a GC content above $0.45$ (empty blue histogram in Fig.~\ref{fig:2}C) tend to have larger characteristic lengths than genomes with a GC content below (empty purple histogram).
After accounting for these differences, however, still only 70\% of the variance of the histograms can be explained by statistical fluctuations from a OU process (filled histograms in Fig.~\ref{fig:2}C). This is consistent with the fact that autocorrelations do not decrease exponentially (Fig.~\ref{fig:3}A). In fact, as we show below, GC correlations occur within specific evolutionary conserved segments.

Finally, a finer analysis reveals differences between the autocorrelation functions defined from the first, second and third base of gene codons. Most importantly, we find that the autocorrelation function over the third base recapitulates most of the autocorrelation computed over all bases (Fig.~\ref{fig:3}B). For distances larger than 5 kb, the autocorrelation function over the first base is typically 4 fold smaller (green curve), and the autocorrelation function over the second base 10 fold smaller (blue curve). These trends are also apparent in individual genomes: $d_{\text{GC}}$ correlates poorly with the characteristic lengths computed from the first and second bases of codons (\suppFig\ref{fig:S2}A-B) but better with the characteristic length computed from the third base (\suppFig\ref{fig:S2}C).

In summary, our analysis indicates the existence of a nearly universal characteristic length of 10 to 20 kb associated with the GC content of bacterial genomes, with systematic variations related to the overall GC content and most of the variability, although not all, imputable to the finite size of chromosomes. It further suggests that the physical phenomenon underlying this characteristic length involves the usage frequency of gene codons. The signal is indeed mostly driven by the correlation of the third site of gene codons, which is known to relate to the usage frequency of synonymous codons~\cite{Wan:2004go}.

\subsection{Characteristic lengths associated with evolutionary conservation}

To identify characteristic lengths, if any, under which genome organization is evolutionary conserved, we analyze finally the conservation of gene clustering along genomes by following and extending our previous approach~\cite{Junier:2016po} (see Methods for details).
The core idea is to compute the fraction of genomes in which a given pair of genes is found within a distance $d^*$ along the chromosome. Here we take $d^*=100$ kb, a distance much larger than any of the characteristic lengths that the following analysis reveals. To compare genes in different genomes, we rely on their classification into orthology classes, i.e., families of phylogenetically and functionally related genes. Given two genes $i,j$ (more precisely, two orthology classes), we first identify the genomes in which genes from these two classes are present and then compute the fraction $f_{ij}$ of those genomes where the genes are within $d^*$. In doing so, we ignore pairs that are represented in less than 100 genomes and account for the possible multiplicity of genes from the same orthology class within the same genome. Following previous works, we also correct for the uneven sampling of available genome sequences by discounting genomes from over-represented clades (see details in Methods).

\begin{figure}[t]
\begin{center}
\includegraphics[width=.66\linewidth]{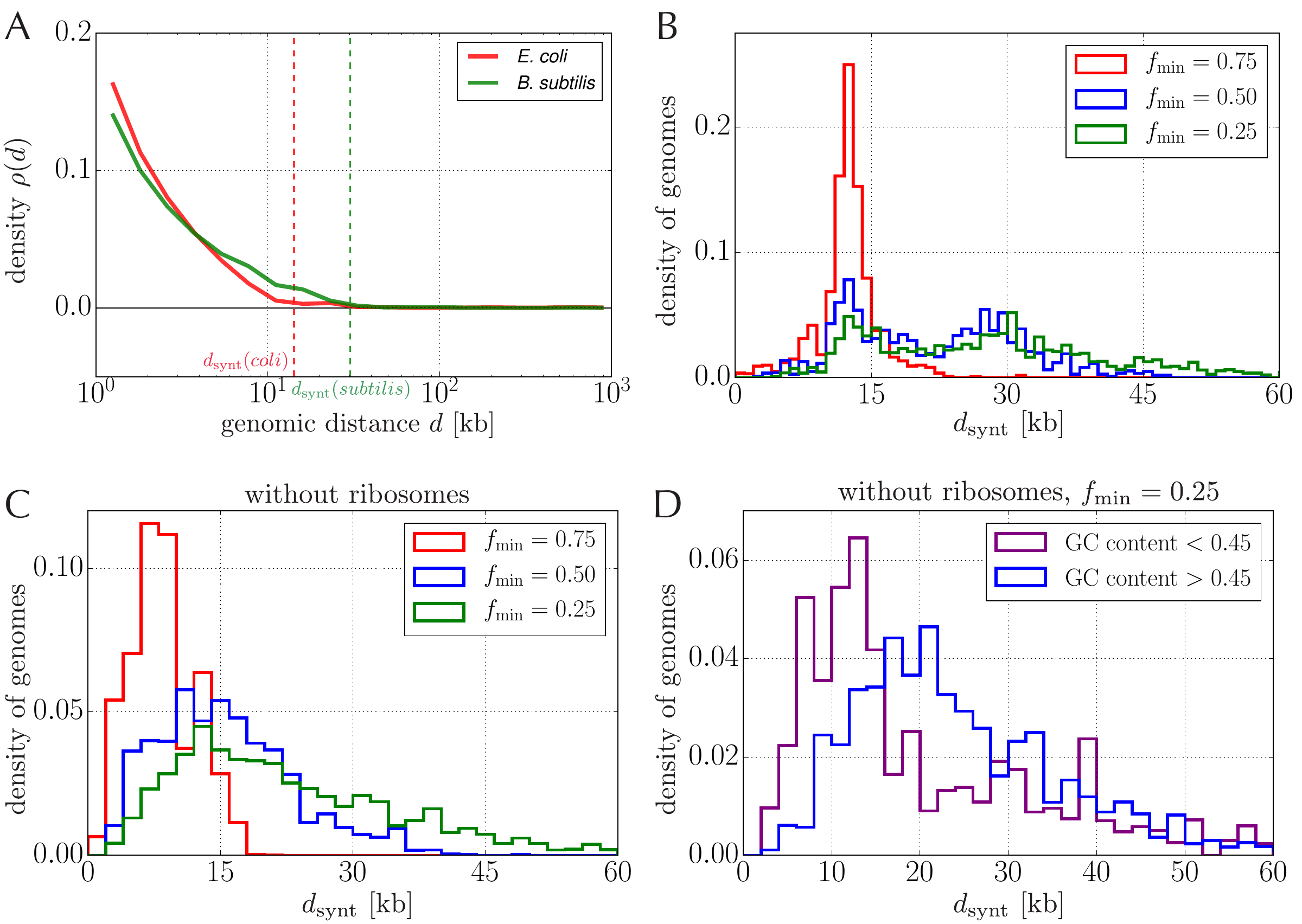}
\caption{Characteristic lengths from evolutionary conservation of gene contexts -- {\bf A.} Distribution $\rho(d)$ of distances between genes found within $d^*=100$ kb of each other in at least a fraction $f_{\rm min}=0.5$ of the genomes of our dataset (in red for {\it E. coli}, in green for {\it B. subtilis}). We extract from these curves characteristic lengths $d_{\rm synt}$ below which correlations are significantly enhanced (dotted lines). {\bf B.} Distribution of characteristic lengths $d_{\rm synt}$ over the 797 chromosomes of the dataset, here computed for three values of $f_{\rm min}$. The distribution is unimodal at the highest level of conservation ($f_{\rm min}=0.75$, in red) but bimodal for lower levels of conservation ($f_{\rm min}=0.5$ and $f_{\rm min}=0.25$, in blue and green), with a first mode around 10 kb and a second around 30 kb. {\bf C.} Ignoring ribosomal genes significantly alters the distribution of characteristic lengths $d_{\rm synt}$ and leaves a single mode around 15 kb. {\bf D.} As in Fig.~\ref{fig:2}C, genomes with high overall GC content (in blue) tend to have larger characteristic lengths than genomes with lower overall GC content (in purple).
\label{fig:4}}
\end{center}
\end{figure}

As a result of our analysis, pairs of genes are sorted by the fraction $f_{ij}$ of genomes in which they are found below $d^*=100$ kb. For each level of conservation $f_{\rm min}$, ``syntenic pairs'' are defined as those satisfying $f_{ij}>f_{\rm min}$. We then analyze within each particular genome the distances $d$ at which the genes associated with these pairs are found. By definition, these distances are below $d^*=100$ kb in a fraction $\geq f_{\rm min}$ of the genomes but, apart from that, the distribution $\rho(d)$ within a given genome of these distances is not constrained by the method. Remarkably, however, the distributions $\rho(d)$ are far from uniform: instead, they are concentrated at short distances, much below $100$ kb; this is illustrated in Fig.~\ref{fig:4}A, which is obtained with the {\it E. coli} and {\it B. subtilis} genomes using $f_{\rm min}=0.5$.

To compare distributions for different values of $f_{\rm min}$ and across different bacteria, we define for each $f_{\rm min}$ and each genome a characteristic length $d_{\text{synt}}$ below which the distribution $\rho(d)$ is significantly enriched. We follow here the same approach as when defining $d_{\text{GC}}$: we consider the mean $m_\rho$ and the standard deviation $\s_\rho$ of the probability density of distances $\rho(d)$ for $d>100$ kb and identify the characteristic length $d_{\text{synt}}$ as the shortest distance where $\rho(d)$ reaches $m_\rho+\s_\rho$. For {\it E. coli} and {\it B. subtilis}, this procedure gives respectively $d_{\text{synt}}= 14$ kb and $d_{\text{synt}}= 30$ kb when considering $f_{\rm min}=0.5$ (dotted lines in Fig.~\ref{fig:4}A). These values, however, vary with the choice of $f_{\rm min}$ (\suppFig\ref{fig:S3}).

Considering all 797 bacterial genomes, we observe that the distribution of characteristic lengths is generically bimodal, with a first mode around 10-15 kb and a second around 30 kb (Fig.~\ref{fig:4}B). As when analyzing correlations in GC content, we must be careful, however, that localized heterogeneities may heavily influence the observations. We find indeed that the 30 kb length results from a particular subset of genes, the subset of ribosomal genes: repeating the analysis without considering them results in unimodal distributions centered around 15 kb, consistent with the characteristic length scale identified from GC contents (Fig.~4C). Further consistency with the analysis of GC content is reported in Fig.~\ref{fig:4}D: similar to Fig.~\ref{fig:2}C, evolutionary characteristic lengths $d_{\text{synt}}$ tend to be larger for genomes with higher overall GC content.

In summary, the evolutionary conservation of gene context leads to essentially the same results as the analysis of GC content: a nearly universal characteristic length around 15 kb, with systematic differences between high and low GC genomes.

\subsection{GC content versus synteny}

\begin{figure}[t]
\begin{center}
\includegraphics[width=.75\linewidth]{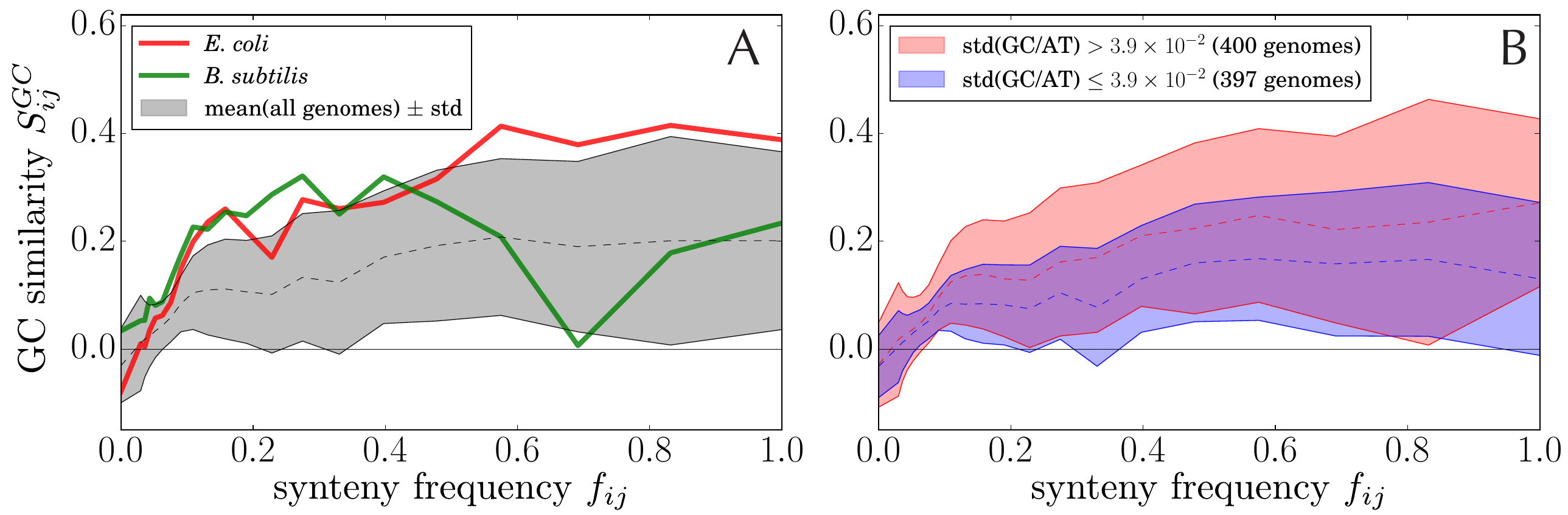}
\caption{Comparison between evolutionary conservation of gene contexts (synteny) and GC content -- {\bf A.} Comparison between the synteny frequency $f_{ij}$ of a pair $ij$ of genes (the fraction of genomes in which they are found within $d^*=100$ kb) and a measure $S^{\text{GC}}_{ij}$ of GC content similarity between the two genes defined to have zero mean when averaged over all pairs of genes (\mm). The relation between the two quantities is shown in red for {\it E. coli}~and in green for {\it B. subtilis}, with the mean over all genomes reported by the dotted line and the standard deviation by the shaded area. Pairs of genes in synteny ($f_{ij}>0.05$) are found to have a similar GC content. {\bf B.} As in A, but separately for the bacterial genomes with largest (in red) and smallest (in blue) variance in GC content. We observe that the similarity of GC content is higher for pairs of genes in synteny in genomes with largest variations in GC content.
\label{fig:5}}
\end{center}
\end{figure}

The convergence between the analyses of GC content and synteny strongly suggests that they reflect the same underlying property of genomes. To compare the two results more directly, we define a similarity in GC content $S^{\text{GC}}_{ij}$ for each pair $ij$ of genes (\mm) and represent it as a function of the frequency of synteny $f_{ij}$ of the pair (Fig.~\ref{fig:5}A). As expected, for \ecoli\ (red curve) and \bsub\ (green curve), and more generally for all 797 studied genes, the two quantities are correlated: pairs of genes with low synteny frequency (e.g. $f_{ij}<0.05$) show poor or no similarity in GC content ($S^{\text{GC}}_{ij}\simeq 0$) while pairs with significant synteny clearly do. Note here that a synteny frequency $f_{ij}=0.1$ is already very significant and that only $\sim 10\%$ of the studied gene pairs have a synteny frequency $f_{ij}>0.1$.

A finer analysis reveals that similarity of GC content between genes in synteny is all the stronger that the variance of the GC content of the genome is larger (Fig.~\ref{fig:5}B). The top 400 genomes with largest variance (red area in Fig.~\ref{fig:5}B) include in particular the GC-homogeneous genome \ecoli, indicating that this trend is not necessarily a consequence of localized heterogeneities such as observed in the genome of \bsub.

\section{Discussion}

From an analysis of hundreds of bacterial genomes, we conclude that GC content correlations, defined from the sequences of individual genomes, and genes in synteny, defined from a comparison of gene neighborhoods across multiple genomes, are characterized by a similar characteristic length around 10 to 20 kb. As per our previous work on synteny, this characteristic length can be attributed to segments of DNA that typically encompass several operons and are co-transcribed by facilitation mechanisms that do not require transcription factors~\cite{Junier:2016po}. Our finding that GC correlations mostly arise from the third site of codons further suggests that gene expression within a segment is coordinated at the translational level (since the third site of codons controls the usage frequency of synonymous codons~\cite{Wan:2004go}). Another non-exclusive possibility may also be that since it is less constrained than the other bases, the third base of gene codons encodes specific DNA structural properties. In any case, this reinforces the conclusion that synteny segments are fundamental units of genomes underlying the basal coordination of gene expression. %, at the levels of transcription and translation.

The segments appear to be universally shared across bacterial genomes. In some bacterial chromosomes, however, they co-exist with a lesser number of longer idiosyncratic genomic domains that extend up to a few hundreds kilobases. These are typically prophages that have been acquired by horizontal transfer. As such, they display very distinct GC contents and, consistent with the general conclusion that variations in GC content underly the independent expression of different functional domains, are in some case transcribed by specific RNA polymerases~\cite{krupovic2011genomics}. Segments of intermediate length around $ 70$ kb associated with horizontal transfers have also been reported for closely related strains of {\it E. coli} ~\cite{Pang:2016gk}. These segments are, however, far less conserved than synteny segments discussed in this work and, although they have distinct GC contents~\cite{papanikolaou2009gene}, are closer in GC content to the rest of the genome than prophages of \bsub.

The presence of large but localized fluctuations in GC contents required us to revisit the application of autocorrelation functions to the identification of characteristic lengths in genomes. Autocorrelation functions are indeed well suited for translationally invariant systems, as typically  encountered in condensed matter physics. In inhomogeneous systems such as genomes, however, large localized heterogeneities can dominate the correlations and conceal more widespread but smaller patterns of correlations. The issue also arises when studying synteny segments, which seem at first sight to display not one but two characteristic lengths. However, the largest length around 30 kb originates from a very small subset of genes. In contrast, the presence of a fundamental and universal characteristic length around 10-20 kb emerges as a statistically robust finding. 

This conclusion leads to a simple question: What ultimately sets the characteristic length of 10 to 20 kb? In particular, is it stemming from the physics of DNA or from the biology of gene regulation? Physically, the segments are consistent with the structuring of bacterial chromosomes into supercoiled domains, but no fundamental constraint is known that limits the length of these domains. Biologically, they may reflect limitations on transcriptional or translational regulation but again, no fundamental constraint is known that limits the distance at which gene expression can be coordinated.

\appendix
\section*{METHODS}
\label{app:meth}

\subsection{Datasets}
{\bf Genomes --} Our input is a collection of complete and well-annotated genomes whose genes are assigned to orthology classes. We rely on data from the STRING database~\cite{STRING} which classifies genes into $C=4866$ clusters of orthologous genes (COGs). As of December 2017, this database covers 2031 taxons, including 1675 bacterial strains. Within the bacterial genomes, we selected those containing a chromosome of length $> 1$ Mb and where at least 60\% of the genes are assigned to COGs. The former constraint was used to avoid artefacts in the identification of fundamental length scales (all much below 1 Mb), while the latter constraint was used to mitigate noise in our evolutionary analysis. Our analysis was based on the $M=797$ chromosomes resulting from this selection (when multiple chromosomes satisfying these criteria were present we took the largest one). The list of these chromosomes is provided in \url{File S1}. We verified that the disregarded genomes have correlations in GC content similar to the selected ones (\suppFig\ref{fig:S1}).

\subsection{Genomic analysis}

{\bf GC content autocorrelations on partionned genomes --} 
Autocorrelation functions are sensitive to the presence of localized motifs. To circumvent this problem, we partition a genome into subsequences of length $L$, compute an autocorrelation function $\Gamma_j(d)$ for each subsequence $j$ and take the median at each given value of $d$: $\Gmed(d)=\text{med}(\{\Gamma_j(d)\}_j)$. In Fig.~\ref{fig:2} , we consider subsequences of length $L=\SI{500}{kb}$.

Contrary to full genomes, we do not apply periodic boundary conditions to the subsequences. Instead, for a subsequence starting at bin $i_0$ and ending at bin $i_1$, where the bin length is $\ell$ $(=\SI{1}{kb})$, we define for each distance $d=0,\ell,2\ell,\dots, (N-1)\ell$ the $d$-shifted signal $y_{d}[i]= x[i+d/\ell]$, for every $i$ in $[i_0,i_1-d/\ell]$. Respectively calling $m_{y,d}$ and $\s_{y,d}$ the mean and standard deviation of $y_d$, the autocorrelation function $\Gamma_{[i_0,i_1]}(d)$ over the subsequence $[i_0,i_1]$ is then defined as
\beq
\Gamma_{[i_0,i_1]}(d)=\frac{1}{(i_1-i_0-d/\ell+1)\s_{y,0}\s_{y,d}}\sum_{i=i_0}^{i_1-d} (y_0[i]-m_{y,0})(y_d[i]-m_{y,d}).
\label{eq:gamma}
\eeq\\

{\bf GC similarity --} To compare GC correlation with synteny properties, we define a measure of GC similarity between two genes. We start with the definition
\beq
s^{\text{GC}}_{ij}=1-\frac{|x_i-x_j|}{|x_i+x_j|}
\eeq
where $x_i$ and $x_j$ are respectively the GC contents of genes $i$ and $j$. We then define a zero-centered measure of GC similarity as
\beq
S^{\text{GC}}_{ij}=
\frac{s^{\text{GC}}_{ij}-\langle s^{\text{GC}}_{ij}\rangle}{\max s^{\text{GC}}_{ij}-\langle s^{\text{GC}}_{ij}\rangle}
\eeq
where $\max s^{\text{GC}}_{ij}$ and $\langle s^{\text{GC}}_{ij}\rangle$ are respectively the maximal and average value of $S^{\text{GC}}_{ij}$ over all pairs $ij$ of genes.\\

\subsection{Evolutionary analysis}
{\bf Genome weights --} From the $M=797$ genomes annotated with $C=4866$ COGs, an $M\times C$ occurrence matrix $O$ is defined by $O_{si}=1$ if strain $s$ contains at least one instance of COG $i$, and 0 otherwise. From this matrix, we define the similarity between strains $s$ and $t$ as 
\beq
S_{st}=\frac{\sum_i O_{si}O_{ti}}{\max_{s,t}(\sum_iO_{si},\sum_i O_{si})}.
\eeq
The quantity $\Delta_{st}=1-S_{st}$ may be interpreted as a phylogenetic distance. Given a threshold $\delta$, we then assign to each strain $s$ a statistical weight $w_s$ defined by 
\beq\label{eq:w}
w_s^{-1} = \sum_t\1[\Delta_{st}<\delta],
\eeq
where the sum is over the strains $s$ and where $\1[X]$ is a generic indicator function with $\1[X]=1$ if and only if $X$ is true. In words, the weight of $s$ is inversely proportional to the number of strains $t$ at phylogenetic distance $1-S_{st}<\delta$; as those strains include $s$ itself, $0<w_s\leq 1$.

On this scale, the distance between {\it E. coli} and {\it B. subtilis} is $\Delta=0.43$, close to the mean distance between all strains $\langle\Delta\rangle=0.47$, and the mean distance between the 8 strains of {\it E. coli} present in our dataset is $\Delta=0.06$ with a maximum at $\Delta=0.1$, which also corresponds to the mean distance between {\it E. coli} and {\it Salmonella} strains. We take this value, $\delta=0.1$, as a cut-off in Eq.~\eqref{eq:w}. This corresponds to an effective number of genomes $M'=\sum_sw_s=553$ (we verified that taking $\delta=0.2$ does not alter our conclusions).\\%, as shown in Fig S1.\\

{\bf Co-occurence --} We use these weights when taking averages over genomes from different strains. In particular, we estimate the effective number of genomes where COG $i$ and $j$ co-occur as
\beq
M_{ij}=\sum_s w_s\1[i\cap s\neq\emptyset]\1[j\cap s\neq\emptyset].
\eeq
 Here, $\1[i\cap s\neq\emptyset]=1$ if $i$ is represented in strain $s$ and 0 otherwise.\\
 
 {\bf Synteny --} Given two COGs $i,j$, we estimate the effective number of strains in which $i$ and $j$ are both present and within a given distance $d^*$ as
\beq
X_{ij}=\sum_s w_s\1[d_{ij}<d^*],
\eeq
where $d_{ij}$ is the distance between genes along a chromosome measured in base pairs. We take $d^*=100$ kb, chosen to be much larger than the characteristic lengths that we find. More generally, to account for the possible presence within a same genome of multiple pairs of genes in two given COGs $ij$, we correct Eq.~\eqref{eq:X} by averaging over all these pairs:
\beq\label{eq:X}
X_{ij}=\sum_sw_s\frac{1}{| i\cap s|| j\cap s|}\sum_{g_i\in i\cap s, g_j\in j\cap s}\1[d_{g_ig_j}<d^*],
\eeq
where $i\cap s$ is as before the set of genes in COG $i$ and in strain $s$ and $|i\cap s|$ the size of this set. Finally, the frequency $f_{ij}$ reporting the fraction of genomes in which $i$ and $j$ at distance $d_{ij}<d^*$ is defined as
\beq
f_{ij}=\frac{X_{ij}}{M_{ij}}.
\eeq
For these fractions to be significant we restrict our analysis to the pairs $ij$ that satisfy $M_{ij}>100$. In {\it E. coli} , this corresponds to 2122 pairs involving 2687 different genes and in {\it B. subtilis} to 2122 pairs involving 2372 genes.

Our formula are identical to those used in our previous work~\cite{schober2017evolutionary} with only one difference: in Eq.~\eqref{eq:X}, we use a common distance $d^*=100$ kb for all strains rather than a strain-specific distance $d_s^*=2p^*/L_s$ that varies with the length $L_s$ of the chromosome. In principle, the later choice is required to define a null model where every pair of genes has same probability $p^*$ to be found in synteny in all genomes. In this work, however, we fix $d^*$ to make clear the fact that the value of $d^*$ has no incidence on the characteristic lengths that we find. In practice, the two definitions lead to indistinguishable results since the characteristic lengths are much shorter than $d^*$ or any of the $d_s^*$.

\newpage

% Hack for making figures Say \figurename S\thefigure, e.g. Figure S1:
\makeatletter
\makeatletter \renewcommand{\fnum@figure}
{\figurename~S\thefigure}
\makeatother
\setcounter{figure}{0}

\section*{SUPPLEMENTARY FILES}

\url{FileS1} lists the 797 chromosomes used in this study. Each line provides, separated by ``:'', (1) the taxon id, (2) the fll name of the strain, (3) the chromosome id; (4) the length of the chromosome (in bp).

\section*{SUPPLEMENTARY FIGURES}

\begin{figure}[h]
\begin{center}
\includegraphics[width=\linewidth]{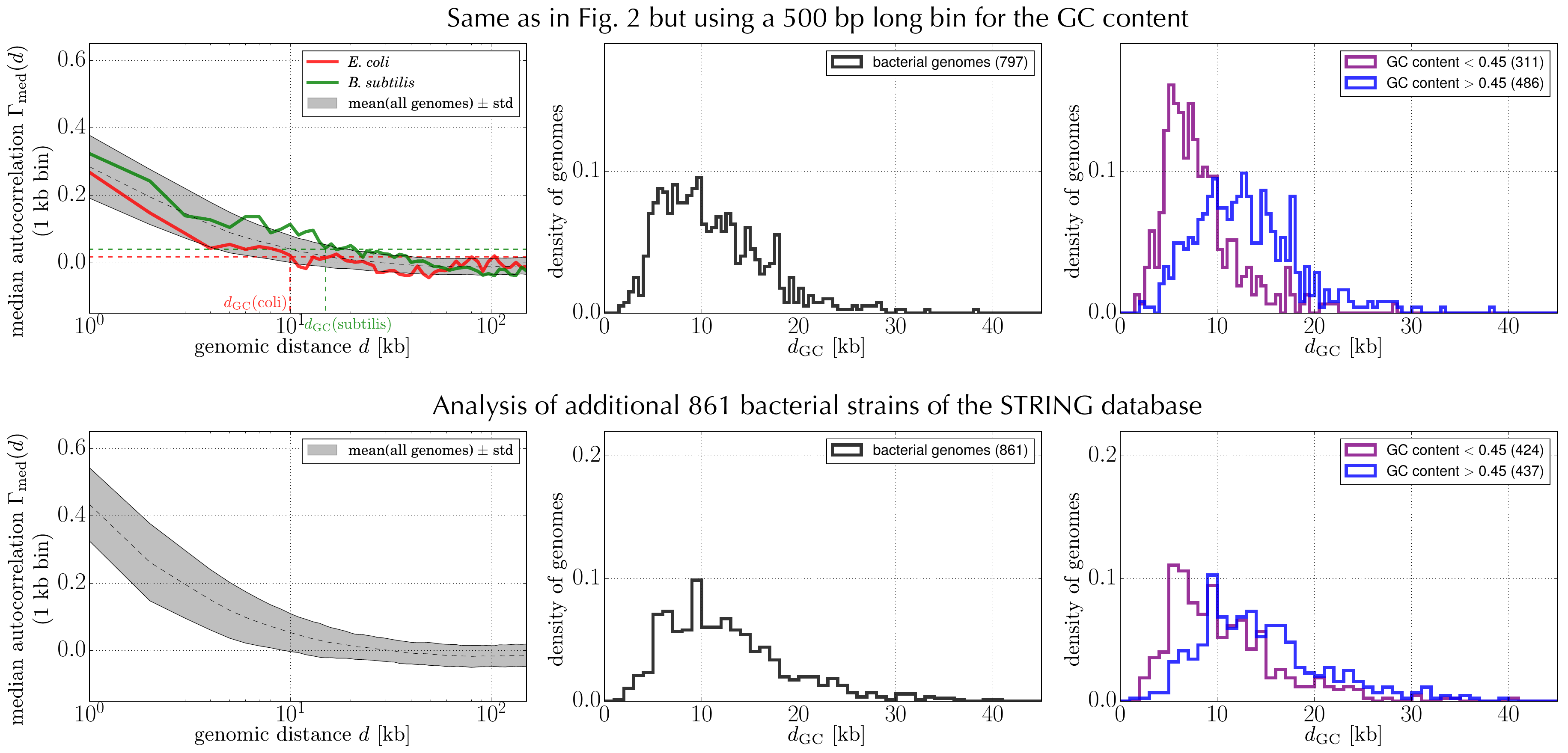}
\caption{Panels on the first line: same as Fig.~\ref{fig:2} (without the OU analysis) but considering bins of length $\ell=500$ bp instead of $\ell=1000$ bp to compute GC contents, showing that characteristic lengths are similar with shorter bins. Panels on the second line: same as Fig.~\ref{fig:2} (without the OU analysis) but considering the 861 genomes of length $\geq$ 500 kb from the STRING database that we discarded in the main figures because of their unsuitability for the evolutionary analysis (see Methods).\label{fig:S1}}
\end{center}
\end{figure}

\vspace{1cm}

\begin{figure}[h]
\begin{center}
\includegraphics[width=.85\linewidth]{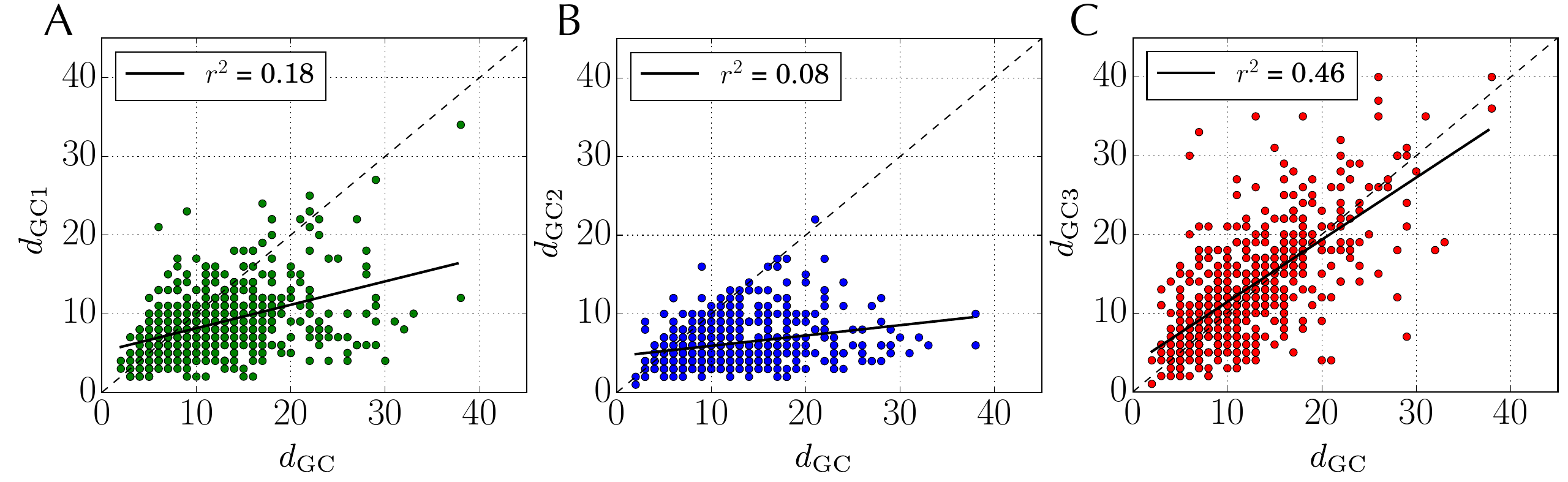}
\caption{Characteristic lengths $d_{\text{GC1}}$, $d_{\text{GC2}}$ and $d_{\text{GC3}}$ computed respectively from the first, second and third base of codons, as a function of the characteristic length $d_{\text{GC}}$ of the overall GC signal, showing that the highest correlation is obtained with the third base.\label{fig:S2}}
\end{center}
\end{figure}

\begin{figure}[h]
\begin{center}
\includegraphics[width=.4\linewidth]{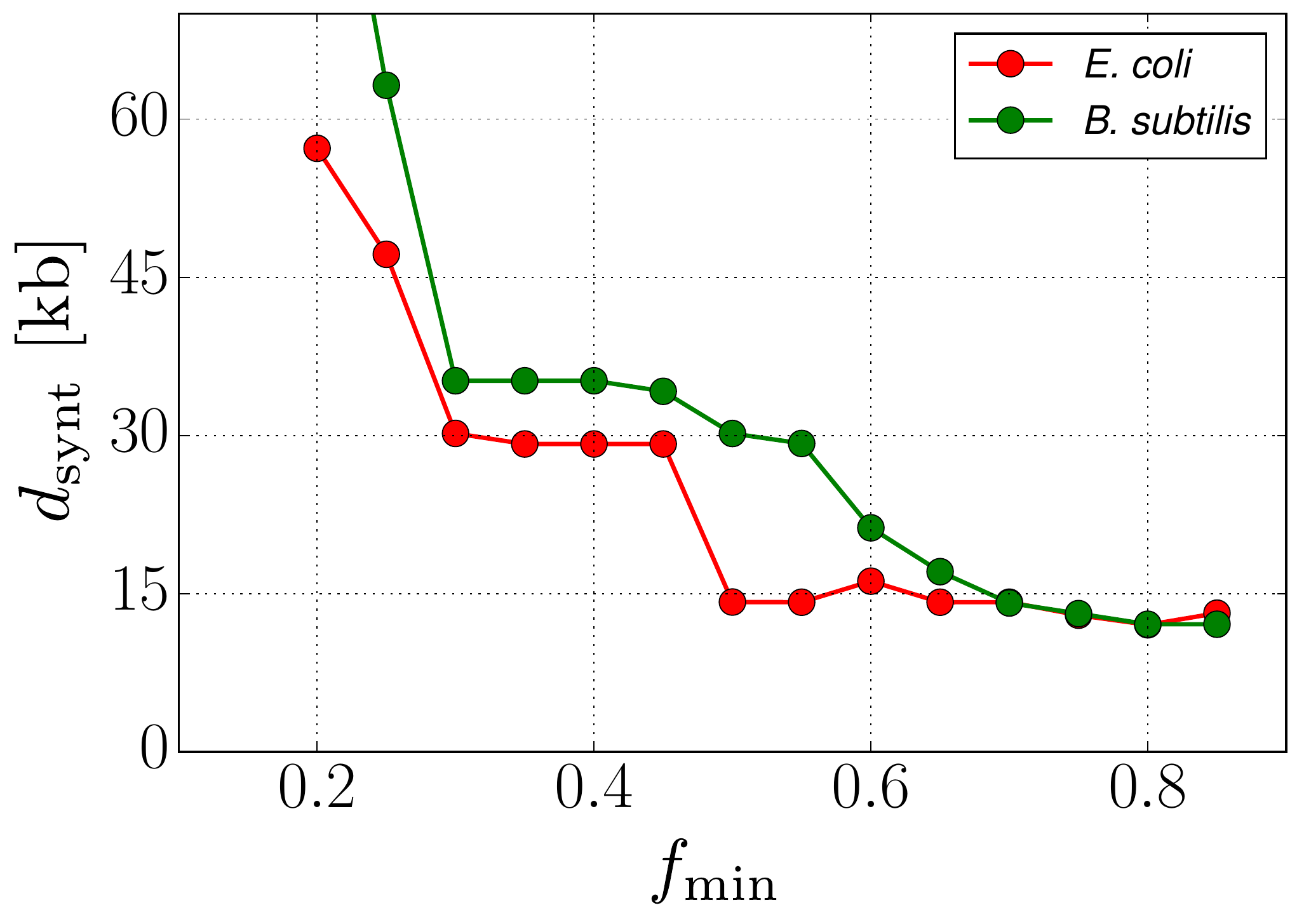}
\caption{Characteristic lengths from evolutionary conservation of gene context. As in Fig.~\ref{fig:4}, but showing here how the characteristic lengths $d_{\rm synt}$ vary as a function of $f_{\rm min}$, the frequency above which two genes are considered in synteny, for a given genome (the genome of {\it E. coli} in red and of {\it B. subtilis} in green). The two plateaus around 15 kb and 30 kb, which are particularly stricking for {\it E. coli} (red), show that the two modes of Fig.~\ref{fig:4}B can also be observed within a given genome.\label{fig:S3}}
\end{center}
\end{figure}

\end{document}